\documentclass[pra,aps,twocolumn]{revtex4-2} %{revtex4-2}
\usepackage{graphicx,amsmath,physics,color,soul}

\usepackage[colorlinks=true,linkcolor=blue,citecolor=blue,urlcolor=black]{hyperref}

\graphicspath{{Images/}}

\begin{document}

\title{Ground-State Cooling of a Mechanical Oscillator via a Hybrid Electro-Optomechanical System}

\author{Roson Nongthombam} 
\email{n.roson@iitg.ac.in}
\author{Ambaresh Sahoo}
\email{ambareshs@iitg.ac.in}
\author{Amarendra K. Sarma}
\email{aksarma@iitg.ac.in}
\affiliation{Department of Physics, Indian Institute of Technology Guwahati, Assam 781039, India}
\date{\today}

\begin{abstract}

We present a scheme for ground-state cooling of a mechanical resonator by simultaneously coupling it to a superconducting qubit and a cavity field. The Hamiltonian describing the hybrid system dynamics is systematically derived. The cooling process is driven by a red-detuned ac drive on the qubit and a laser drive on the optomechanical cavity. We have investigated cooling in the weak and the strong coupling regimes for both the individual system, i.e., qubit assisted cooling and optomechanical cooling, and compared them with the effective hybrid cooling. It is shown that hybrid cooling is more effective compared to the individual cooling mechanisms, and could be applied in both the resolved and the unresolved sideband regimes.

\end{abstract}

\maketitle

%--------------------------------------
\section{Introduction} \label{intro}
%--------------------------------------
\noindent
The realization of a macroscopic mechanical oscillator in the quantum regime has a wide range of applications in studying fundamental physics and developing quantum technologies ranging from high-precision measurements to quantum information processing \cite{Aspelmeyer14,Liu13-2,sarma2021continuous,Kirill}. For this, the mechanical oscillator is cooled down to its quantum ground state. This can be done by first cryogenically pre-cooled to about a few thousand initial phonons and then further cool down to the ground state by coupling to external dissipation sources. There are two particular sources of dissipation that are studied extensively, both theoretically and experimentally. One source is contributed from interaction with a cavity field in an optomechanical system, and the other one by coupling with a Cooper-pair box (CPB) qubit.

Cavity optomechanics is the study of light interacting with a harmonically bound movable mirror placed inside a cavity \cite{Aspelmeyer14,Bowen_book,Liu13,Liu13-2,Kippenberg08}. A basic cavity optomechanical experiment was first conducted by Braginsky and co-worker in the year 1967 \cite{Braginsky67,Braginsky70}, using a microwave cavity and later in the optical domain by Dorsel and his team, 1983 \cite{Dorsel83}. The first experimental radiation-pressure cooling of a mechanical oscillation using optical feedback was demonstrated by Cohedon, Heidmann, and Pinard (1999) \cite{Mancini98,Cohadon99}, and later cooled down to much lower temperature using the same approach \cite{Kleckner06,Corbitt07,Poggio07}. In 2006, radiation pressure cooling of a micromechanical resonator down to an effective temperature of 10K was realized \cite{Gigan06,Arcizet06}. Cooling in the resolved sideband regime was achieved in 2008 \cite{Schliesser08}. 
Cooling close to the ground state under cryogenically pre-cooled environment was later demonstrated \cite{Groblacher09,Park09,Schliesser09}. The other method of cooling a mechanical resonator using a superconducting qubit is theoretically studied in Refs.\cite{Martin04,Hauss08,Jaehne08,Grajcar08}. It has been demonstrated in Refs.\cite{Regal08,Teufel11} that it is possible to integrate a nanomechanical resonator into a superconducting transmission line microwave cavity. Coupling of a Cooper pair box superconducting qubit with an on-chip superconducting transmission line resonator is demonstrated in Refs.\cite{Wallraff04,Blais04,Koch07,Blais20}.  Capacitive coupling of a micromechanical resonator and a Cooper pair box qubit is realized in Ref.\cite{Pirkkalainen13}.  Despite the above experimental advancements, the qubit-assisted ground-state cooling of a resonator is less known. In this context, studies of hybrid systems comprising of optomechanical and electromechanical systems have got a lot of attention \cite{chu2020,Vitali12,xiang2013hybrid,Forsch,Tang21}. Recently a detailed theoretical study of ground-state cooling of a radio frequency (rf) resonator using an optoelectromechanical system formed by an optical cavity, a mechanical oscillator and a MHz rf resonator is reported in Ref.\cite{Vitali21}.

In this work, we study the ground state cooling of a mechanical resonator by simultaneously coupling the resonator to an optical cavity field via radiation pressure and to a CPB qubit via a movable capacitive plate. A similar system where the resonator is piezoelectrically coupled with the qubit is demonstrated in Ref.\cite{Mirhosseini20}. It is also shown that a microwave signal can be efficiently converted to an optical one \cite{Andrews14,Barzanjeh11}. It should be noted that cooling of  the mechanical oscillator either optomechanically \cite{Mancini98,Cohadon99,Kleckner06,Corbitt07,Poggio07,Gigan06,Arcizet06, Schliesser08,Groblacher09,Park09,Schliesser09,Marquardt08,Marquardt07} or using superconducting qubit \cite{Martin04,Hauss08,Jaehne08,Grajcar08} has been studied previously by many researchers. However, in this work, we propose a scheme where the combined effect of these two processes is incorporated.
 We red-detune the qubit and the optical cavity for cooling by applying an external ac drive and shining with bright coherent light, respectively.  We find that the coupling between the qubit and the resonator is affected due to the coherent drive. We compare the individual cooling with the hybrid one for weak and strong coupling. In the weak coupling condition, we study cooling in the resolved and unresolved sideband regime. For specific choices of parameters, we find more cooling in the hybrid case.
 
The rest of the paper is organized as follows: In section\,\ref{Sec2}, we systematically derive the Hamiltonian of the hybrid system. Then, we discuss the cooling scheme for both the weak and the strong coupling regime in section\,\ref{Sec3}. Concluding remarks are given in section\,\ref{Sec4}. Finally, two Appendices are presented to display the detailed calculations related to section\,\ref{Sec2}.  

%------------------------------------
\begin{figure}[t]
\centering
\begin{center}
\includegraphics[width=0.45\textwidth]{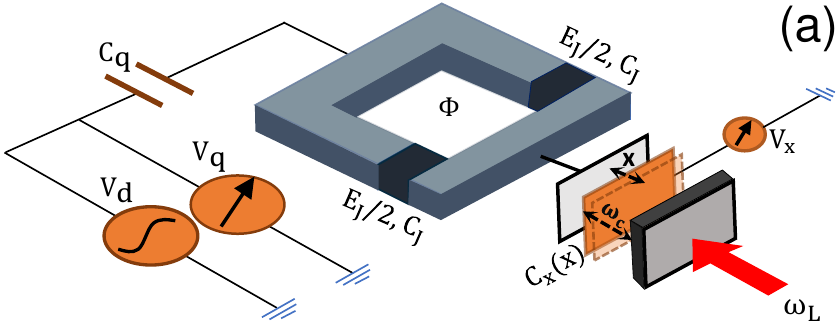}\\
\vspace{0.5cm}
\includegraphics[width=0.26\textwidth]{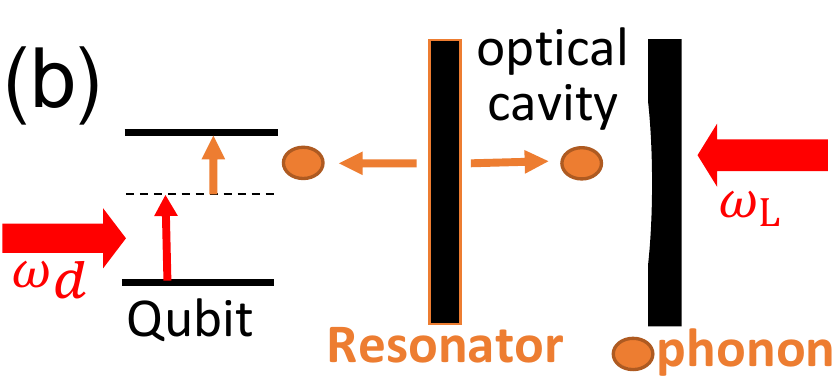}
\caption{(Color online) Hybrid CPB qubit and optomechanical system. (a) Schematic description of the hybrid system under study. A split CPB electrostatically biased by voltages $V_q$ and $V_x$ and driven by voltage $V_d$ is capacitively coupled to a mechanical mode via movable capacitor $C_x(x)$. By applying external flux $\Phi$ and changing the charging energy $E_c$, the energy levels of the CPB qubit can be adjusted. The mechanical mode is coupled to an optical cavity field $(\omega_c)$ formed by placing a partially reflecting mirror in front of the movable capacitor plate. The optical cavity is driven by a bright coherent light. (b) Schematic description of the cooling process. The qubit and the optical cavity are red-detuned. The qubit gets excited by absorbing incoming photons from the external drive and phonons from the mechanical oscillator. The optical photons enter the cavity by absorbing a phonon from the oscillator.}  
\label{fig:schem}
\end{center}
\end{figure}
%--------------------------------------

%--------------------------------------
\section{Hybrid System} \label{Sec2}
%--------------------------------------
\noindent
We consider a hybrid system comprising of a Josephson junction superconducting qubit (CPB qubit) and an optomechanical system as shown schematically in Fig.\,\ref{fig:schem}(a). This hybrid system consists of an intermediary mechanical oscillator, coupled to a pair of Josephson junction qubits through the capacitance $C_x(x)$ and an optical mode.   $C_x(x)$ depends on the mechanical displacement $x$, caused by the radiation pressure interaction (optomechanical coupling) . Adopting the same approach as in Ref.\cite{Martin04} for the Josephson qubit part, the Hamiltonian of the hybrid system without dissipation could be written as follows:
\begin{align} \label{eq1}
\hat{H}= & \frac{\left[\hat{Q}-Q_{xq}(x)\right]^2 }{2C_{\Sigma }(x)}-E_J(\Phi_{\rm ext}) \cos\theta+ \hbar \Omega \, \hat{b}^{\dagger}\hat{b} \nonumber \\ &\hspace{0.2cm}- \hbar \Delta_c \, \hat{a}^{\dagger}\hat{a}
 + \hbar g_o \, \hat{a}^{\dagger}\hat{a}\left(\hat{b}^{\dagger}+\hat{b} \right) + \hbar \eta \left(\hat{a}^{\dagger}+\hat{a}\right).
\end{align}
Here, $\hat{Q}=2e\hat{N}$ (in the number basis), where $\hat{N}$ is the number operator for the Cooper pairs transferred across the Josephson junction in the superconducting qubit (CPB).
$Q_{xq}(x)=2eN_{xq}(x)$, where $N_{xq}(x)=N_x(x) + N_q$, is the offset charge or gate charge produced by external gate voltages  $V_x=2eN_x(x)/C_x(x)$ and $V_q=2eN_q/C_q$, which induces Cooper pairs to tunnel through the Josephson junction, and hence control the charge and state of the CPB. It is apparent that this offset charge is dependent on the resonator displacement $x$. $C_{\Sigma}(x)=2C_J+C_q+C_x(x)$ is the total capacitance of the qubit. $E_J(\Phi_{\rm ext})\cos\theta$ is the effective field in the two parallel junctions, each with energy $E_J/2$. $\Phi_{\rm ext}$ is the external flux applied in the loop formed by the two junctions. Here, $\theta$ is the phase difference between the junctions and $E_J(\Phi_{\rm ext})=E_J \cos(\pi\Phi_{\rm ext}/\Phi_o)$. The movable capacitor in the system acts as a mechanical oscillator with frequency $\Omega$ and is described by the third term in the Hamiltonian. This movable capacitor and a partially reflecting mirror placed in front of it form an optomechanical system. The last three terms in Eq.\,\eqref{eq1} describe the optomechanical system in the driven frame.  Here, $\Delta_c (=\omega_L-\omega_c)$ is the detuning between the laser drive ($\omega_L$) and the cavity ($\omega_c$) frequency. The second-last term describes the coupling between an optical mode ($\hat{a}$) in the cavity and a mechanical mode $(\hat{b})$ via optomechanical coupling rate $g_o$. The last term is the laser drive with amplitude $\eta$.

If the qubit is driven coherently at frequency $\omega_d$  and amplitude $\Omega_R$, the Hamiltonian described by Eq.\,\eqref{eq1}, could be expressed in the qubit basis, subject to certain approximations, as given below (see Appendix\,\ref{AppendixA} for details):
\begin{align} \label{eq2}
	\hat{H}=&- \frac{\hbar\Delta_q}{2}\sigma_z+\frac{1}{2}\hbar\Omega_R\,\sigma_x \cos\varphi+\hbar g\left(\hat{b}^{\dagger}+\hat{b}\right)\sigma_z \sin\varphi \nonumber \\ &+\frac{\hbar g^2}{\omega_q}\left(2\hat{b}^{\dagger}\hat{b}+1\right)\sigma_z \cos^2 \varphi +\hbar \Omega \, \hat{b}^{\dagger}\hat{b} - \hbar \Delta_c \,\hat{a}^{\dagger}\hat{a} \nonumber \\ &
	\hspace{2cm}  +g_o\hbar\hat{a}^{\dagger}\hat{a}\left(\hat{b}^{\dagger}+\hat{b}\right) + \hbar\eta\left(\hat{a}^{\dagger}+\hat{a}\right).
\end{align}
Here, $\Delta_q=\omega_d-\omega_q$, where $\omega_q$ is the transition frequency of the qubit. $\sigma_x$ and $\sigma_z$ are the usual Pauli matrices. $g$ is a coupling constant. The parameter $\varphi$ is defined through $\tan\varphi=\epsilon/E_J$, with $\epsilon$ being associated with the variation of gate charge. For details of the parameters, please refer to Appendix\,\ref{AppendixA}.  

The optomechanical coupling rate $g_o$ is usually smaller than the mechanical ($\gamma$) and optical ($\kappa$) decay rates. One common approach to address this issue is to drive the optical cavity using strong coherent light. This drive significantly increases the radiation pressure force, and hence the optomechanical coupling rate. It also induces a classical steady-state displacement of both the intracavity field and the mechanical mode \cite{Bowen_book}. The quantum fluctuations around the classical steady-state values are small. Hence, we make the following transformation, also referred to as shifted or displaced frame \cite{Aspelmeyer14}: $\hat{a}\rightarrow\alpha+\delta\hat{a}$ and $\hat{b}\rightarrow\beta+\delta\hat{b}$, where $\alpha$ and $\beta$ are respectively, the steady-state displacements of the intracavity field and the mechanical mode, while $\delta\hat{a}$ and $\delta\hat{b}$ are the corresponding quantum fluctuations. Using this transformation in Eq.\,\eqref{eq2} and removing the constant terms, we obtain:
\begin{align} \label{eq3}
&\hspace{-0.2cm}\hat{H}= -\hbar \frac{\Delta_q}{2}\sigma_z+\frac{1}{2}\hbar\Omega_R\,\sigma_x \cos\varphi+\hbar g \left(\hat{b}^{\dagger}+\hat{b}\right)\sigma_z \sin\varphi \nonumber \\
 &\hspace{-0.30cm} +\frac{2 \hbar g^2\beta}{\omega_q}\left(\hat{b}^{\dagger}+\hat{b}\right)\sigma_z \cos^2\varphi+\frac{2 \hbar g^2}{\omega_q}\hat{b}^{\dagger}\hat{b}\sigma_z \cos^2\varphi -\hbar \Delta_c\hat{a}^{\dagger}\hat{a} \nonumber \\ &\hspace{-0.35cm} +\hbar g_o[\alpha(\hat{a}^{\dagger}+\hat{a})+\hat{a}^{\dagger}\hat{a}]\left(\hat{b}^{\dagger}+\hat{b}\right)  +\hbar \Omega\, \hat{b}^{\dagger}\hat{b}+H_a +H_b.
\end{align}
Here, $\hat{a} \Rightarrow \delta\hat{a}$, $\hat{b} \Rightarrow \delta\hat{b}$, $\Delta_q \Rightarrow \Delta_q-({4g^2\beta}/{\omega_q})\cos^2\varphi|\beta|^2-2g\beta \sin\varphi$, and $\Delta_c\Rightarrow\Delta_c-2g_o\beta$. 
The last two terms in Eq.\,\eqref{eq3} constitute the terms that are proportional to $\hat{a}$, $\hat{a}^{\dagger}$, $\hat{b}$ and $\hat{b}^{\dagger}$ (see Appendix\,\ref{AppendixB} for details). The fourth and fifth terms are the qubit and oscillator interaction terms. Two observations could be made when we compare these two interacting terms. Firstly, the fourth term is amplified by the steady-state displacement of the mechanical oscillator $\beta$. Secondly,  the fourth and the fifth term gives rise to second and third-order nonlinearity respectively. The same observations could also be deduced for the seventh term. However, in this case, the coupling is between the optical photon and mechanical phonon, and the second-order nonlinear coupling is amplified by the coherent amplitude $\alpha$. Because the second-order nonlinear interaction terms are amplified, we neglect the third-order interaction. The resultant Hamiltonian could then be put in the following form:    
\begin{align}
&\hspace{-0.2cm}\hat{H}= -\hbar\frac{\Delta_q}{2}\sigma_z+\frac{1}{2}\hbar\Omega_R \, \sigma_x+\hbar G\left(\hat{b}^{\dagger}+\hat{b}\right)\sigma_z  +\hbar \Omega \, \hat{b}^{\dagger}\hat{b} \nonumber \\
  &\hspace{0.5cm} -\hbar \Delta_c \,\hat{a}^{\dagger}\hat{a} +\hbar G_o\left(\hat{a}^{\dagger}+\hat{a}\right)\left(\hat{b}^{\dagger}+\hat{b}\right) +H_a+H_b,
\end{align}
where $G=g\sin\varphi+({2g^2\beta}/{\omega_q})\cos^2\varphi$, $G_o=g_o\alpha$, and $\Omega_R\rightarrow \Omega_R \cos\varphi$. 
We observe that in the presence of the steady-state mechanical displacement $\beta$, induced through the optical drive, the coupling rate between the qubit and the oscillator increases by a factor $({2g^2\beta}/{\omega_q})\cos^2\varphi$. Similarly, the optomechanical coupling rate is amplified to $g_o\alpha$.

%-----------------------------------
\section{Cooling}  \label{Sec3}
%------------------------------------
\noindent
The cooling of the mechanical oscillator, using the hybrid system, could be obtained in the two coupling regimes, namely, the weak and the strong coupling regime. In the weak coupling regime, the qubit and the optical cavity field act as perturbations to the oscillator. However, in the strong coupling regime, we assume that only the qubit acts as a perturbation, and the optical field strongly couples with the oscillator. Apart from its own environment (bath), the mechanical oscillator has two additional sources of dissipation, one from the qubit and the other from the optical resonator. Due to the different timescales of these dissipation channels ($n_{th}\gamma \ll \Gamma, \kappa$), they act separately. Assuming that the timescale of evolution of the bath is much shorter than the timescale for the interaction between the bath and the system, the Lindblad master equation for the hybrid system in the shifted frame reads as follows:
\begin{align} \label{eq16}
\dot{\hat\rho}&=-\frac{i}{\hbar}[\hat{H},\hat{\rho}]+(\mathcal{L}_q+\mathcal{L}_m+\mathcal{L}_c)\hat{\rho} \nonumber
    \\ & \hspace{2.2cm} +\frac{\kappa}{2}\alpha[\hat{a}-\hat{a}^{\dagger}, \hat{\rho}] +\frac{\gamma}{2}\beta[\hat{b}-\hat{b}^{\dagger}, \hat{\rho}],
\end{align}
where \vspace{-0.2cm}
\begin{align}
&\mathcal{L}_q=\Gamma(n_q+1)D[\sigma_{-}](\hat{\rho})+ \Gamma n_qD[\sigma_{+}](\hat{\rho}) +\frac{\Gamma_d}{2}(\sigma_z\hat{\rho}\sigma_z-\hat{\rho}), \nonumber \\
& \mathcal{L}_m=\gamma(n_{th}+1)D[\hat{b}](\hat{\rho})+\gamma n_{th}D[\hat{b}^{\dagger}](\hat{\rho}), ~~ \mathcal{L}_c=\kappa D[\hat{a}](\hat{\rho}), \nonumber \\
& {\rm and}~~ D[\hat{A}]=\frac{1}{2}(2\hat{A}\hat{\rho}\hat{A}^{\dagger}-\hat{A}^{\dagger}\hat{A}\hat{\rho}-\hat{\rho}\hat{A}^{\dagger}\hat{A}).
\end{align}
The decay rates $\Gamma_d$, $\Gamma$ and $\gamma$ can be found out from the noise correlations of gate voltage fluctuations or charge number fluctuations; $\hat{\delta N}_x$ and $\hat{\delta N}_q$. Similarly, the optical cavity decay rate $\kappa$ can be derived from the noise correlation of the optical bath. 

\subsection{Weak coupling}
\begin{figure}[ht]
\centering
\begin{center}
\includegraphics[width=0.235\textwidth]{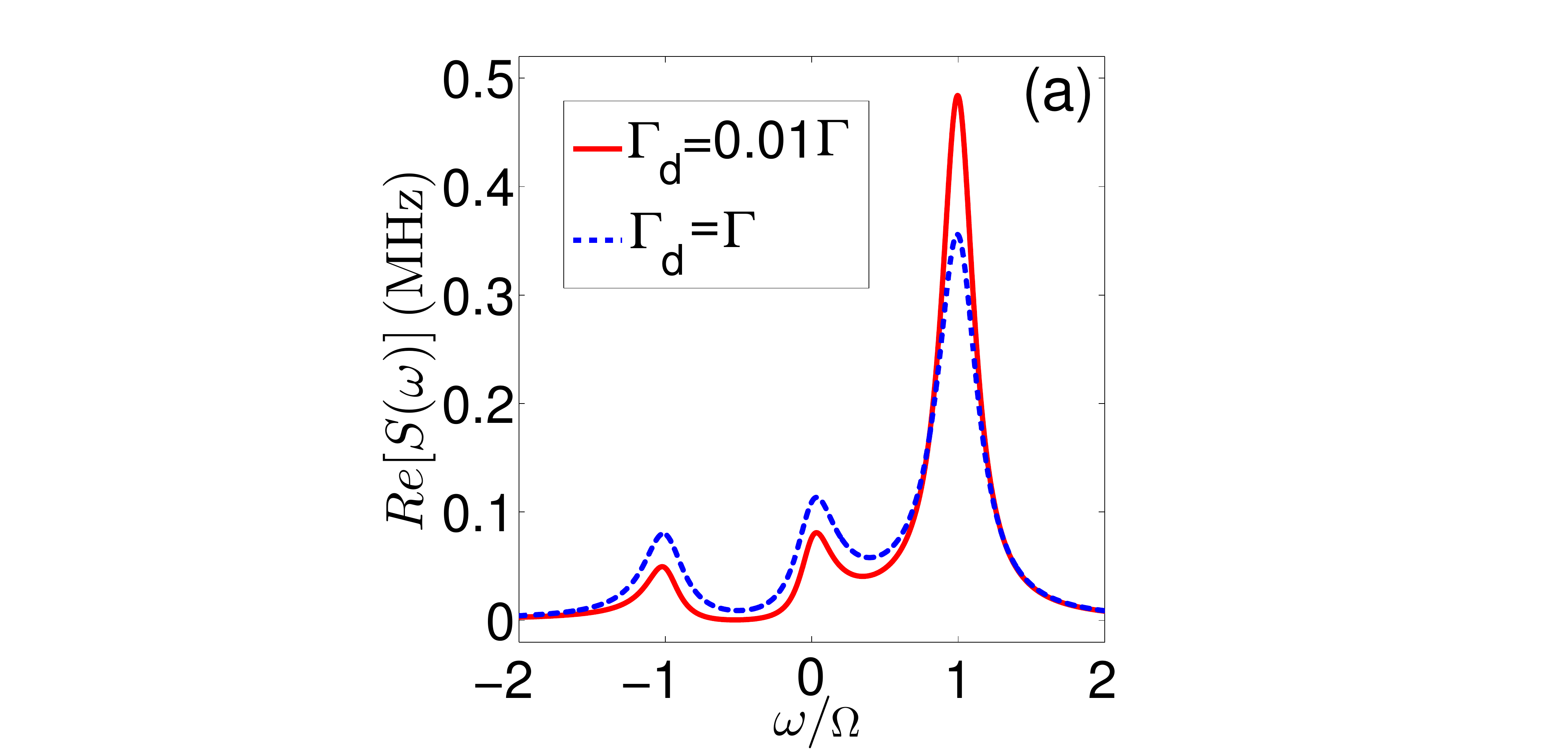}
\includegraphics[width=0.23\textwidth]{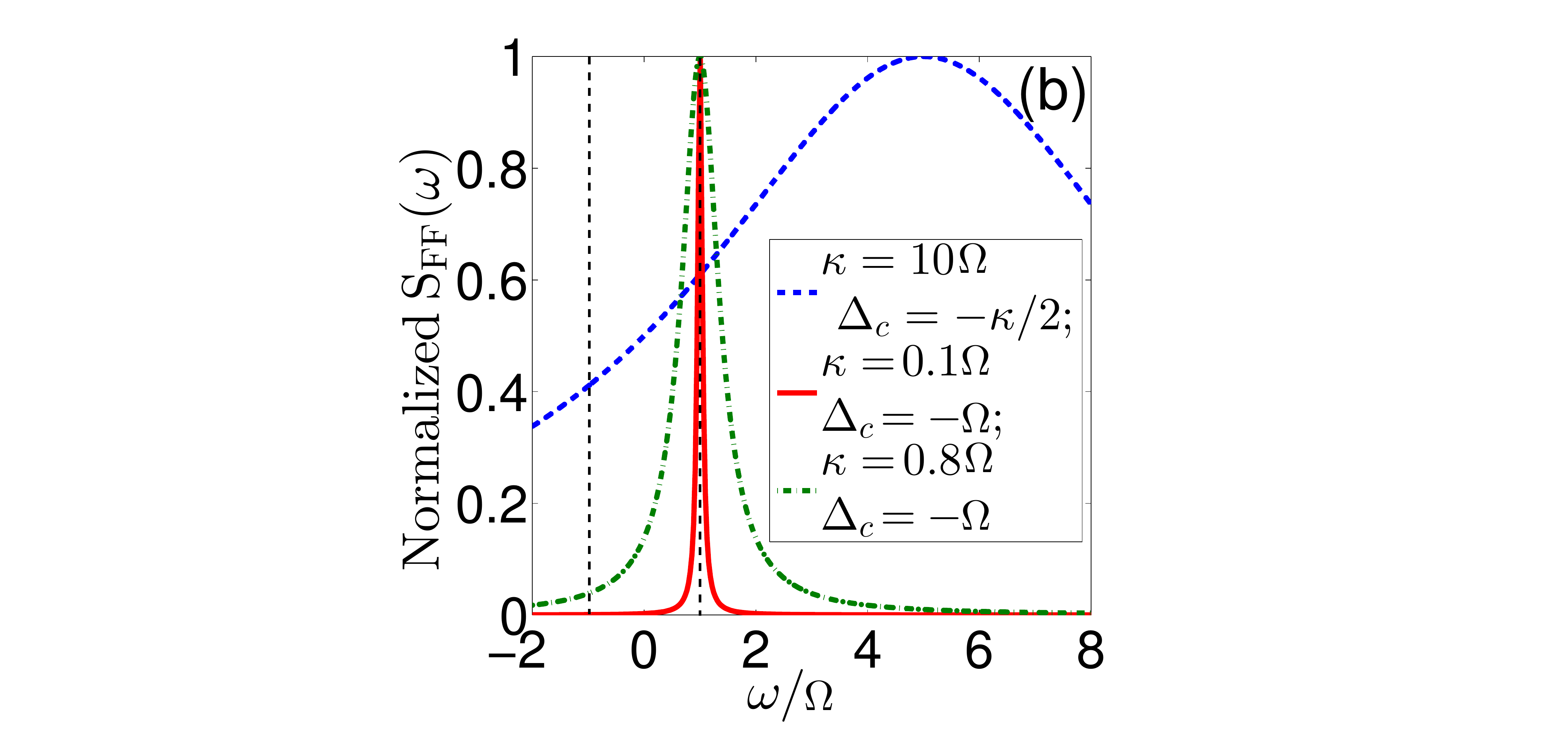}
\caption{ (Color online) (a) Real part of the qubit spectral noise density at optimal cooling drive strength $\Omega_R=0.85\,\Omega$ and resonance frequency $\bar{\Delta}_q=\Omega$. The parameters used for the plot are $\Omega=10$\,MHz and $\Gamma=2$\,MHz. The solid red curve is for the dephasing rate $\Gamma_d=0.01\Gamma$, and the dotted blue curve corresponds to $\Gamma_d=\Gamma$\,MHz. The difference between the peaks at $\pm\Omega$ is more and the heating peak at $\Omega=-10$ is less when $\Gamma \gg \Gamma_d$, and hence more cooling. (b) Spectral noise density of the radiation force in the resolved (solid red and dotted green curve) and unresolved sideband (dotted blue curve) regime. The parameters used here correspond to that of minimum quantum limit cooling. Minimum quantum limit is attained at $\Delta_c=-\kappa/2$ in the unresolved sideband $(\kappa\gg \Omega)$, and at $\Delta_c=-\Omega$ in resolved sideband $(\kappa\ll \Omega)$.}  \label{fig:2}
\end{center}
\end{figure}

The weak coupling regime assumes $G_o \ll \kappa$ and $G \ll \Gamma$, which means that the qubit and the cavity-optical field go to equilibrium before their states could hardly be affected by the mechanical interaction. Thus, the cavity optical field and the qubit can be adiabatically eliminated using the Nakajima-Zwanzig formalism \cite{Jaehne08,Zwanzig64,Wilson-Rae07}. Since the qubit and the cavity field are not coupled, we can eliminate them separately. To eliminate the qubit, we first split the total density operator by means of a projection operator $P$ and $Q$ as
\begin{equation}
\hat{\rho}=(P+Q)\hat{\rho},  ~~~{\rm and } ~~~   P+Q=I,
\end{equation}
with $P$ defined by
\begin{equation}
P\hat{\rho}=\hat{\rho}_{qss}\otimes\hat{\rho}_{om}, \qquad \hat{\rho}_{om}={\rm tr}_{q}[\hat{\rho}],
\end{equation}
where $\hat{\rho}_{qss}$ is the steady-state density operator of the qubit, and $\hat{\rho}_{om}$ is the optomechanical density operator.
Projecting Eq.\,\eqref{eq16} into $P$-space, the master equation reads:
\begin{align}\label{eq9}
P\dot{\hat\rho}=&P[-i\Omega \hat{b}^{\dagger}\hat{b},P\hat{\rho}]-\frac{i}{\hbar}P[H_a+H_b,P\hat{\rho}] \nonumber \\
&  +P[i\Delta_c\hat{a}^{\dagger}\hat{a}-iG_o(\hat{a}^{\dagger}+\hat{a})(\hat{b}^{\dagger}+\hat{b}),P\hat{\rho}] \nonumber \\
&  +P[-iG(\hat{b}^{\dagger}+\hat{b})\sigma_z ,Q\hat{\rho}+P\hat{\rho}] \nonumber \\
&  +P(\mathcal{L}_m+\mathcal{L}_c)P\hat{\rho}+P\frac{\gamma}{2}\beta[\hat{b}-\hat{b}^{\dagger}, P\hat{\rho}]  \nonumber \\
& +P\frac{\kappa}{2}\alpha[\hat{a}-\hat{a}^{\dagger}, P\hat{\rho}].
\end{align}
Similarly, the master equation in the $Q$-space can be obtained by projecting Eq.\,\eqref{eq16} into the $Q$ space. The steady-state displacements $\alpha$ and $\beta$ are determined from Eq.\,\eqref{eq9} as follows:
\begin{align}
\hspace{-0.3cm}\alpha=\frac{\eta}{\Delta_c+i{\kappa}/{2}-g_o(\beta+\beta^*)}, ~
\beta=\frac{g_o|\alpha|^2+G\langle\sigma_z\rangle_s}{i{\gamma}/{2}-\Omega}. 
\end{align}
Here, $\langle\sigma_z\rangle_s=P\sigma_z$ is the steady-state expectation value of $\sigma_z$. 
For $g_o \ll 1$, $\Delta_c \gg \kappa$ and ${\gamma}/{2}\ll\Omega$, we obtain $\beta=-({g_o|\alpha|^2+G\langle\sigma_z\rangle_s})/{\Omega} $ and $\alpha={\eta}/[{\Delta_c-g_o(\beta+\beta^*)}]$ which are real values.
The qubit indirectly interacts with the optical field through the mechanical oscillator. In the following, we assume that the qubit and the optical field evolve independently with no interaction. The master equation for the optomechanical system ($\hat{\rho}_{om}$) can be obtained by solving the coupled rate equations for $P\hat{\rho}$ and $Q\hat{\rho}$ \cite{Jaehne08}:
\begin{align} \label{eq22}
\dot{\hat{\rho}}_{om}=&-\frac{i}{\hbar}[\hat{H}_{om},\hat{\rho}_{om}]+\{\Gamma_q^{+}+\gamma n_{th}\}D[\hat{b}^{\dagger}](\hat{\rho}) \nonumber \\
&  + \{\Gamma_q^{-}+\gamma(n_{th}+1)\}D[\hat{b}](\hat{\rho})+\kappa D[\hat{a}](\hat{\rho}) ,
\end{align}
where
\begin{align}
&\hat{H}_{om}=\hbar \Omega'\, \hat{b}^{\dagger}\hat{b}-\hbar \Delta_c\,\hat{a}^{\dagger}\hat{a} +\hbar G_o\left(\hat{a}^{\dagger}+\hat{a}\right)\left(\hat{b}^{\dagger}+\hat{b}\right), \nonumber \\
 &\Gamma_q^{-}=2G^2Re\{S(\Omega)\},~ ~ \Gamma_q^{+}=2G^2Re\{S(-\Omega)\}, \nonumber \\
 & {\rm and}~~ \Omega'=\Omega+Im\{S(\Omega)+S(-\Omega)\}.
\end{align}
The master equation [Eq.\,\eqref{eq22}] is correct only up to the \textit{$\mathcal{O}$}\{$(G/\Gamma)^2,(\gamma\, n_m/\Gamma)^2$\}. We see that by eliminating the qubit dynamics, the frequency of the oscillator is coherently shifted, and two additional decay rates appear ($\Gamma_q^-$ and $\Gamma_q^+$). The decay rate $\Gamma_q^+$ is responsible for heating, whereas $\Gamma_q^-$ contributes to cooling. The half-sided spectral noise density $S(\omega)$ is defined as
\begin{equation} \label{eq24}
S(\omega)=\int_0^{\infty}dt\,{\rm e}^{i\omega t}\langle \Delta\sigma_z(t)\Delta\sigma_z(0)\rangle,
\end{equation}
where $\Delta\sigma_z=\sigma_z-\langle\sigma_z\rangle_s$. An approximate analytical expression for $Re\{S(\omega)\}$ near resonance condition $\sqrt{\Delta_q^2+\Omega_R^2}=\Omega$ is derived in Ref.\cite{Jaehne08}. Fig.\,\ref{fig:2}(a) shows the plot of $Re\{S(\omega)\}$ under red-detuning ($\Delta_q < 0$) and resonance condition.  Two plots are shown in the figure, one for $\Gamma \gg \Gamma_d$, and the other for $\Gamma\approx\Gamma_d$. As shown in the figure, the difference between the peaks at $\pm \Omega$ is more when the relaxation rate $\Gamma$ is much greater than the dephasing rate $\Gamma_d$. As we will see later in Eq.\,\eqref{eq30}, this difference should be significant for achieving optimal cooling.
%-------------------------------------------------
%------------------------------------------------------

The master equation [Eq.\,\eqref{eq22}] contains the dynamics of both the mechanical oscillator and cavity optical field. The cavity field can be traced out and adiabatically eliminated using the same $P$ and $Q$ formalism 
\begin{equation}
P\hat{\rho}_{om}=\hat{\rho}_{oss}\otimes\hat{\rho}_{m},  \qquad \hat{\rho}_{m}={\rm tr}_{o}[\hat{\rho}],
\end{equation}
where $\hat{\rho}_{oss}$ is the steady-state density operator of the optical field, and $\hat{\rho}_{m}$ is the mechanical density operator.
Analogous with the qubit elimination result, the elimination of the cavity field coherently shifts the mechanical frequency and adds two decay rates ($\Gamma_o^-$ and $\Gamma_o^+$).
The reduced master equation of the mechanical resonator in the interaction picture reads \cite{Wilson-Rae07}:
\begin{align} \label{eq26}
\dot{\hat{\rho}}_{m}=& \{\Gamma_o^{+}+\Gamma_q^{+}+\gamma n_{th}\}D[\hat{b}^{\dagger}](\hat{\rho}) \nonumber \\
& \hspace{1cm}+ \{\Gamma_q^{-}+\Gamma_q^{-}+\gamma(n_{th}+1)\}D[\hat{b}](\hat{\rho}),
\end{align}
with
\begin{align} \label{eq27}
\Gamma_o^{-}&=\frac{x_{\rm ZPF}^2}{\hbar^2}S_{\rm FF}(\Omega)~ ~ {\rm and}~
 \Gamma_o^{+}=\frac{x_{\rm ZPF}^2}{\hbar^2}S_{\rm FF}(-\Omega).
\end{align}
Here $S_{\rm FF}(\omega)$ is the spectral noise density of the radiative force $\hat{F}\,[=(\hbar G/x_{\rm ZPF})(\hat{a}+\hat{a}^{\dagger})]$ acting on the mechanical oscillator:
\begin{equation}
    S_{\rm FF}(\omega)=\int_{-\infty}^{\infty}dt \, {\rm e}^{i\omega t}\langle\hat{F}(t)\hat{F}(0)\rangle.
\end{equation} 
In Fig.\,\ref{fig:2}(b), we plot $S_{\rm FF}(\omega)$ for the resolved ($\kappa=0.1\,\Omega$, $\kappa=0.8\,\Omega$) and unresolved ($\kappa=10\,\Omega$) sideband in the red-detuning case. 
The expectation value of the phonon occupation number can be calculated from Eq.\,\eqref{eq26} as
\begin{equation}
\langle\dot{\hat{n}}_m\rangle=-(\gamma+\Gamma_o+\Gamma_q)\langle\hat{n}_m\rangle+\gamma n_{th}+\Gamma_o^++\Gamma_q^- ,
\end{equation}
where $\Gamma_q=\Gamma_q^- -\Gamma_q^+$ and $\Gamma_o=\Gamma_o^- -\Gamma_o^+$ are the cooling rates contributed from the qubit and optical cavity field, respectively, with
\begin{align}  \label{eq30}
\Gamma_q = & 2G^2\left[{\rm Re}\{S(\Omega)\}-{\rm Re}\{S(-\Omega)\}\right],  \nonumber \\
\Gamma_o= &\frac{x_{\rm ZPF}^2}{\hbar^2}[S_{\rm FF}(\Omega)-S_{\rm FF}(-\Omega)].
\end{align}
For effective cooling, the cooling rates [Eq.\,\eqref{eq30}] must be positive. This condition is met if the external drive of the qubit and optical cavity are red-detuned with respect to the qubit energy level and optical resonance, respectively (see Fig.\,\ref{fig:2}).  This means that the incoming photons from the external drive enter the optical cavity by absorbing a phonon from the mechanical oscillator. In the qubit case, the energy from the incoming photon is less than the excitation energy, and thus energy is taken from the resonator phonon to excite fully. Therefore, in both the cases, phonons are emitted, thereby cooling the mechanical oscillator. However, there is also a finite probability for phonon absorption, thus heating the oscillator. The phonons are absorbed at the rate $\Gamma_q^++\Gamma_o^+$. Cooling happens when the emission rate is  faster than the absorption rate, which is generally the case in the red-detuning. The dynamics at the red-detuning are schematically shown in Fig.\,\ref{fig:schem}(b). The steady-state phonon number of the mechanical oscillator is given by
\begin{equation} \label{eq31}
\langle\hat{n}_m\rangle_{ss}=\frac{\gamma \, n_{th}+\Gamma_o^++\Gamma_q^+}{\gamma+\Gamma_o+\Gamma_q}.
\end{equation}
It is clear from Eq.\,\eqref{eq31} that cooling is prominent when the cooling rates $\Gamma_q$ and $\Gamma_o$ are maximum, and the heating rates $\Gamma_q^+$ and $\Gamma_o^+$ are minimum. The quantum limit for the bare qubit cooling is given by
\begin{equation}
\langle\hat{n}_m\rangle_q={\Gamma_q^+}/{\Gamma_q}.
\end{equation}
The minimum value of this quantum limit turns out to be zero at drive strength $\Omega_R=0$. However, with no drive, the qubit is neither heating nor cooling the oscillator since the spectral noise density responsible for these processes is zero for all frequencies. In the following, we find the minimum quantum limit in the presence of drive.
Using Eq.\,\eqref{eq30} and Eq.\,\eqref{eq24}, we derive the decay rate $\Gamma_q$ for $\Gamma \gg \Gamma_d$ and $\bar{\Delta}_q=\Omega$, as given below:
\begin{equation}
 \Gamma_q=\beta f\left({\Omega_R}/{\bar{\Delta}_q}\right),
\end{equation}
where $\beta=2G^2/\Gamma$,  $\bar{\Delta}_q=\sqrt{\Omega_R^2+\Delta_c^2}$, and 
\begin{equation}
 f\left({\Omega_R}/{\bar{\Delta}_q}\right)=4\frac{\left({\Omega_R}/{\bar{\Delta}_q}\right)^2\sqrt{1-\left({\Omega_R}/{\bar{\Delta}_q}\right)^2}}{4-\left({\Omega_R}/{\bar{\Delta}_q}\right)^4}.
\end{equation}
The maximum cooling rate is achieved for a large value of $\beta$ and the value of $\Omega_R$ that maximize the function $f({\Omega_R}/{\bar{\Delta}_c})$, i.e., $\Omega_R=0.85\,\Omega$. At the optimal drive strength, $\Omega_R=0.85\,\Omega$, the detuning is not absolute, i.e., $\Delta_q=-0.53\,\Omega$. 
The spectral noise density ${\rm Re}\{S(\omega)\}$ at the optimal drive is plotted in Fig.\,\ref{fig:2}(a). The minimum quantum limit for the optimal drive strength is $\langle\hat{n}_m\rangle_{q,min}=0.106$.

The optomechanical cooling rate $\Gamma_o$ could be expressed using Eq.\,\eqref{eq30}, as follows:
\begin{equation}\label{eq25}
\Gamma_o=|G_0|^2\left\{\frac{\kappa}{(\Omega+\Delta_c)^2+{\kappa^2}/{4}} -\frac{\kappa}{(-\Omega+\Delta_c)^2+{\kappa^2}/{4}} \right\}.
\end{equation}
The quantum limit of the cooling induced by the optical cavity field is given by 
\begin{equation}
\langle\hat{n}_m\rangle_o={\Gamma_o^+}/{\Gamma_o},
\end{equation}
which we obtain using Eq.\,\eqref{eq27} and Eq.\,\eqref{eq25} as
\begin{equation}
\langle\hat{n}_m\rangle_o=-\frac{4(\Omega+\Delta_c)^2+\kappa^2}{16\,\Omega\,\Delta_c}.
\end{equation}
The minimum cooling limit $\langle\hat{n}_m\rangle_{o,min}$ is reached at a detuning $\Delta_c=-\sqrt{\Omega^2+\kappa^2/4}$ as
\begin{equation}
\langle\hat{n}_m\rangle_{o,min}=\frac{1}{2}\left(\sqrt{1+\left(\frac{\kappa}{2\Omega}\right)^2}-1\right).
\end{equation}
For slow oscillators, referred to as the unresolved sideband regime, $\Omega\ll \kappa$, the minimum cooling limit is $\langle\hat{n}_m\rangle_{o,min}=\kappa/(4\Omega)$ for $\Delta_c=-\kappa/2$. Ground-state cooling is not possible in this regime. The minimum cooling for fast oscillators, or the resolved sideband regime, $\kappa \ll \Omega$, is $\langle\hat{n}_m\rangle_{o,min}=\kappa^2/(4\Omega)^2$ for $\Delta_c=-\Omega$. Thus, ground-state cooling is possible for the high-frequency oscillator.
The corresponding spectral noise density $S_{\rm FF}(\omega)$ for the minimum cooling limit is shown in Fig.\,\ref{fig:2}(b). 

\begin{figure}[t]
\centering
\begin{center}
\includegraphics[width=0.2315\textwidth]{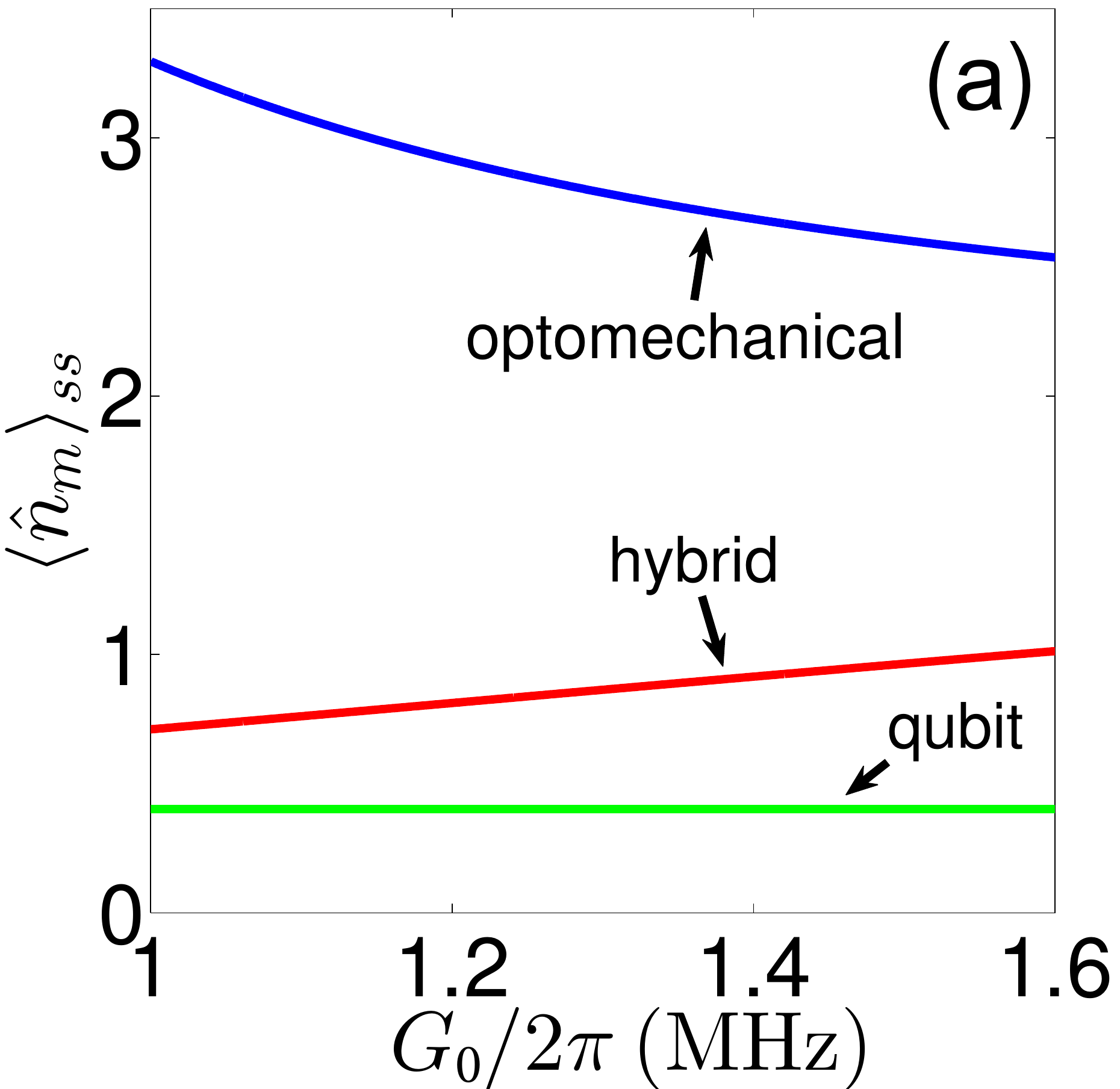}
\includegraphics[width=0.24\textwidth]{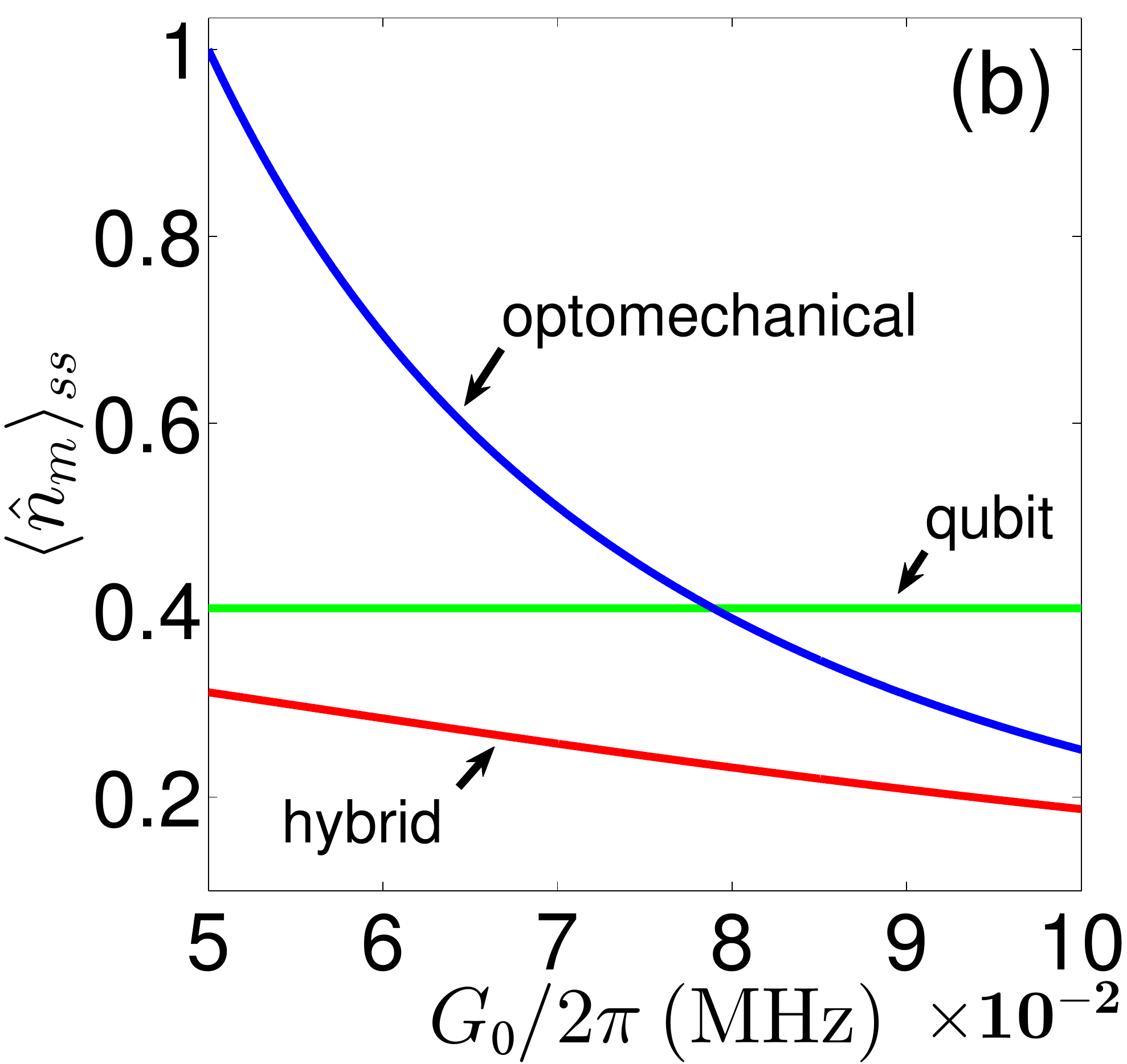}
\caption{(Color online) (a) Steady-state mean phonon number in the unresolved sideband regime,  $\kappa=10\,\Omega$ and $\Delta_c=-\kappa/2$. Green, blue and red curves represent qubit, optomechanical and hybrid  cooling, respectively. Ground state cooling is possible, but at the expense of bare qubit cooling. 
(b) Cooling in the resolved sideband regime, $\kappa=0.1\,\Omega$ and $\Delta_c=-\Omega$. More effective cooling is observed in the hybrid case. Other parameters: $\Omega=10$\,MHz, $G=0.2$\,MHz, $\gamma=10^{-5}$\,MHz, and $n_{th}=10^3$.}  \label{fig:3}
\end{center}
\end{figure}

The effective minimum cooling, $\langle\hat{n}_m\rangle_{ss}$ [Eq.\,\eqref{eq31}], for different optomechanical coupling rates $G_0$ and the qubit coupling rate $G=0.2$\,MHz is shown in Fig.\,\ref{fig:3}(a) and \ref{fig:3}(b).
We see that the qubit brings the mechanical oscillator to its ground state even in the unresolved sideband regime. However, the bare qubit cooling is more effective than hybrid cooling in this regime. Nevertheless, we can study the applications of the optomechanical system in this regime. In the case of a high-frequency oscillator, or resolved sideband regime, we find that hybrid cooling is more efficient than the bare individual coolings.
%
%-----------------------------------------------------
%-----------------------------------------------------------

\subsection{Strong coupling}
\begin{figure}[t]
\centering
\begin{center}
\includegraphics[width=0.235\textwidth]{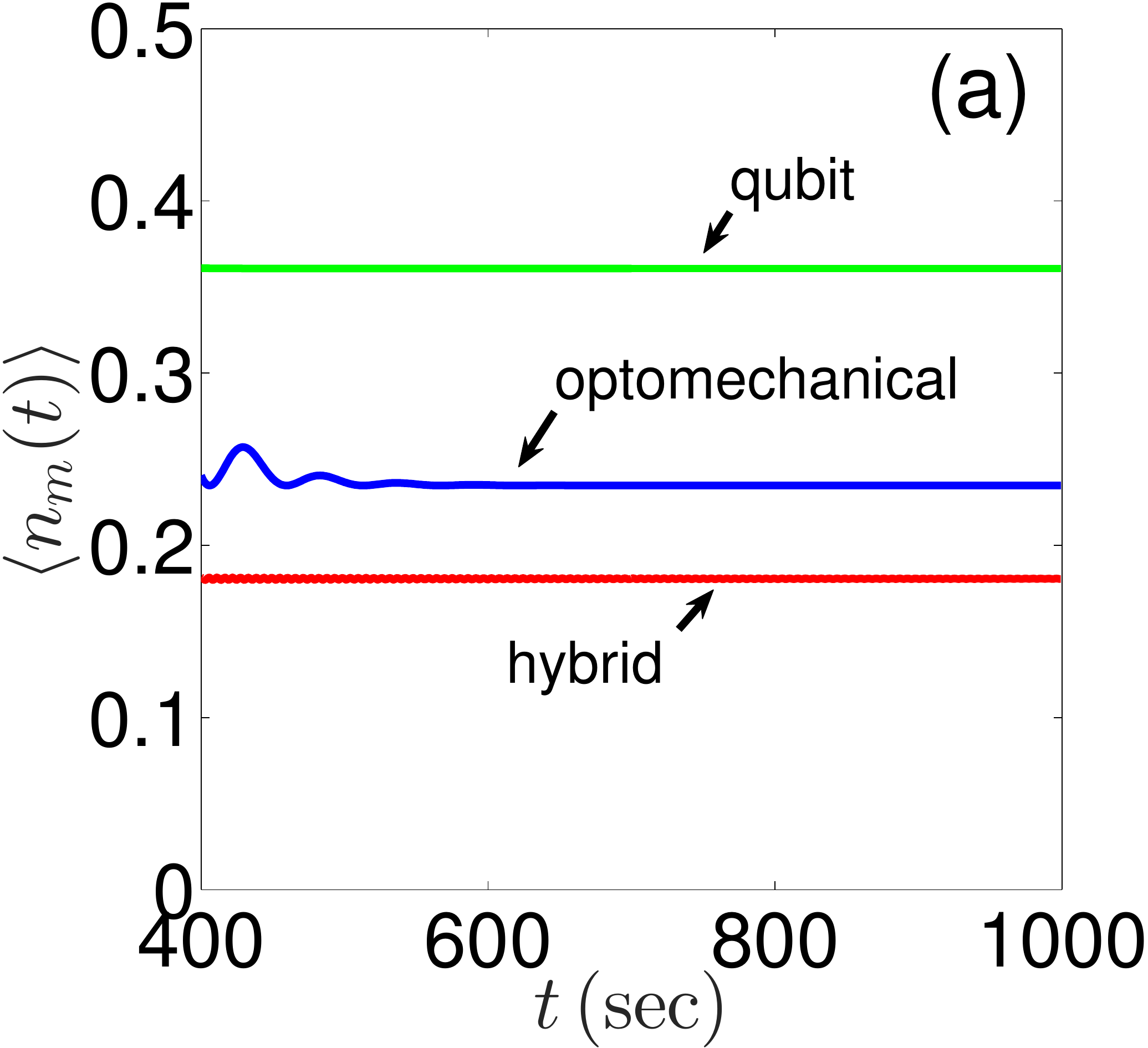}
\includegraphics[width=0.235\textwidth]{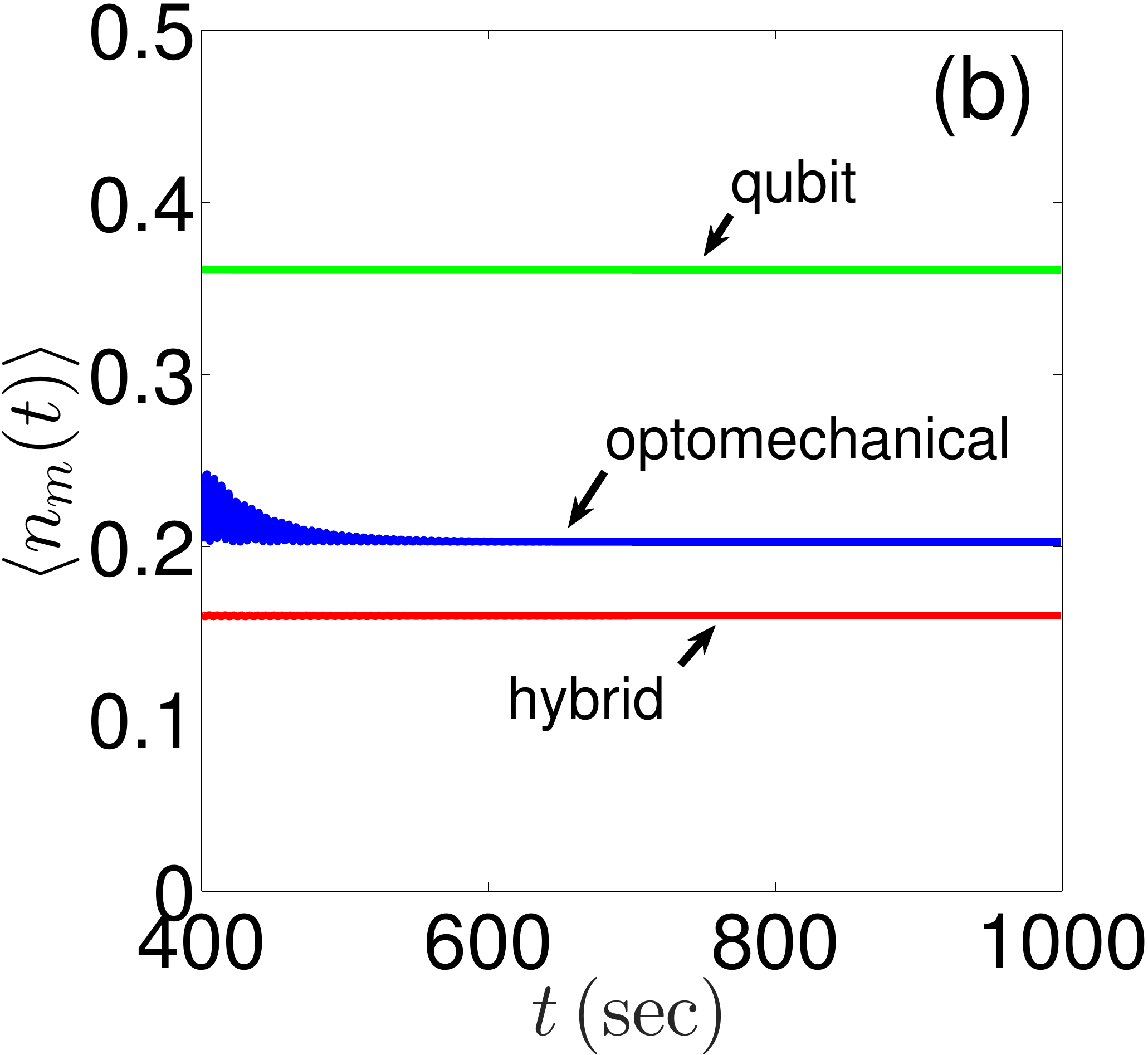}
\caption{(Color online) Time evolution of mean phonon number $\langle\hat{n}_m\rangle(t)$ for (a) $G_o=0.06$\,MHz, $\kappa=0.05$\,MHz, and $G=0.2$\,MHz, and (b) $G_o=0.6$\,MHz, $\kappa=0.05$\,MHz, and $G=0.2$\,MHz. Other parameters: $\Omega=10$\,MHz, $\gamma=10^{-5}$\,MHz, and $n_{th}=10^3$. }  \label{fig:4}
\end{center}
\end{figure}

The strong coupling here refers to the coupling between the optical cavity field and the mechanical oscillator. We assume that the qubit state is hardly affected by the mechanical interaction and goes to steady-state quickly, hence adiabatically eliminated from the dynamics of the hybrid system.
The reduced master equation of the resultant optomechanical system is given by Eq.\,\eqref{eq22}. For determining the steady-state phonon occupation number, we have to solve a linear system of differential equations involving all the the second-order moments: $\langle\hat{a}^{\dagger}\hat{a}\rangle$,  $\langle\hat{b}^{\dagger}\hat{b}\rangle$,
$\langle\hat{a}^{\dagger}\hat{b}\rangle$,  $\langle\hat{a}\hat{b}\rangle$,  $\langle\hat{a}^2\rangle$, and
$\langle\hat{b}^2\rangle$ \cite{Liu13}. The time evolution of the mean phonon number for two coupling cases, $G_o=1.2\,\kappa$ and $G_o=12\,\kappa$, taking the initial phonon numbers to be $10^3$, is shown in Fig.\,\ref{fig:4}.
%-----------------------------------------------------------
%----------------------------------------------------------
It could be seen that in both the coupling cases (intermediate coupling $G_o=1.2\kappa$ and strong coupling $G_o=12\kappa$), the hybrid cooling is more effective than the individual coolings. Furthermore, cooling is more in the strong coupling case than the intermediate coupling one.

\section{Conclusion}  \label{Sec4}
\noindent
We have studied the cooling of a mechanical oscillator in a hybrid system of optomechanics and a superconducting qubit. The Hamiltonian describing the system dynamics is systematically derived. We show that because of the external laser drive, the coupling rate between the qubit and the oscillator is modified. We discussed the cooling effect in two specific regimes, namely, the weak and the strong coupling regimes. In the weak coupling regime, the dynamics of the qubit and the optical cavity field are adiabatically eliminated, which resulted in adding an extra factor in the steady-state displacement of the resonator. The significance of this extra factor is apparent while dealing with optomechanical bistability. We also discussed the quantum limit of cooling for both the individual qubit and optomechanical cooling under the weak coupling regime. We show that in the weak coupling and resolved sideband regime, cooling is more efficient in the hybrid case for a specific choice of parameters. In the unresolved sideband regime, the ground-state cooling of the mechanical resonator is still possible, however, at the expense of qubit cooling. Cooling in the strong optomechanical coupling case is also studied. It is found that hybrid cooling is more effective compared to the individual cooling mechanisms. Finally, we would like to emphasize that the hybrid system proposed in this work could also be used for quantum state transferring purposes and could be utilized as a quantum transducer.

\section*{Acknowledgments}

Roson Nongthombam gratefully acknowledges support of a research fellowship from CSIR, Govt. of India

\appendix
%--------------------------------------------
\section{Derivation of the Hamiltonian [Eq.\,\eqref{eq2}]} \label{AppendixA}
%--------------------------------------------
\noindent In the following, we derive the Hamiltonian [Eq.\,\eqref{eq2}] for the Hybrid system in the qubit basis. Let us begin with Eq.\,\eqref{eq1}. At the gate charge close to an odd number of electron charges, i.e., $N_{xq}(x)=N_x(x) + N_q + 1/2 - \Delta N $, where $|\Delta N |\ll 1/2$,  the first two energy levels of the Josephson junction are very close compared to the higher ones. Therefore, we can make a two-level approximation in this region. To have a clear idea of this approximation, we rewrite the first and second terms of the Eq.\,\eqref{eq1} in the number basis. Denoting the resultant transformation of these two terms by $H_q$, we have
\begin{align} \label{A1}
	H_{q}=&4E_c(x)\sum_N{\left[\hat{N}-N_{xq}(x)\right]^2}\ket{N}\bra{N} \nonumber \\
	&-\frac{E_J}{2}\sum_N{\left(\ket{N+1}\bra{N}+\ket{N}\bra{N+1}\right)},
\end{align}
where $E_c(x)=e^2/2C_{\Sigma}(x)$ is the charging energy of the CPB. Substituting the value of $N_{xq}(x)$, which is close to an odd number of electron charges, in Eq.\,\eqref{A1}, and restricting $\ket{N}$ to $\ket{0}$ and $\ket{1}$, we get the following Hamiltonian:
\begin{align} \label{A2}
	H_q=&-E_c(x)\left[1+4\Delta N(x)^2\right]I \nonumber \\&\hspace{1.5cm} +4E_c(x)\Delta N(x)\sigma_z'- \frac{E_J}{2}\sigma_x'
\end{align}
Here, the first term simply adds extra energy of $E_c(x)(1+4\Delta^2)$ to all the states. Thus, we can omit this energy offset term without loss of any generality, and we get 
\begin{align} \label{A3}
	H_q=-4E_c(x)\Delta N(x)\sigma_z'-\frac{E_J}{2}\sigma_x'.
\end{align}
Assuming that the displacement $x$ of the movable capacitor is relatively small compared to it's initial separation $d$, we obtain $C_x(x)\approx C_x-C_x(x/d)$. Hence $\Delta N(x)\approx\Delta N-N_x(x/d)$ and $E_c(x)\approx E_c+E_c(C_x/C_{\Sigma})(x/d)$. Here, $C_x=2eN_x/V_x$, $E_c=e^2/2C_{\Sigma}$ and $C_{\Sigma}=2C_J+C_q+C_x$.
Then, from Eq.\,\eqref{A3}, we get
\begin{align} \label{A4}
	\hat{H}_q=-\frac{\epsilon}{2}\sigma'_z-\frac{E_J}{2}\sigma'_x-\hbar g\left(\hat{b}^{\dagger}+\hat{b}\right)\sigma'_z,
\end{align}
where $g=(4E_cX_{\rm ZPF}/\hbar d)[N_x-\Delta N(C_x/C_{\Sigma})]$, $\epsilon=8E_c\Delta N$, and $X_{\rm ZPF}$ is the zero-point field amplitude of the oscillator. For a large value of $N_x\gg 1$, the coupling rate could be put as, $g\approx 4E_cX_{\rm ZPF}/(\hbar d N_x)$. Both the gate voltage $V_q$ and $V_x$ have quantum fluctuations ($\hat{\delta N}_x$ and $\hat{\delta N}_q$) which acts as a quantum dissipation to both the qubit and oscillator. These fluctuation are incorporated in $\Delta N$, i.e., $\Delta N\rightarrow \Delta N+\hat{\delta N}_x+\hat{\delta N}_q$.  Substituting these fluctuations in Eq.\,\eqref{A4}, and in the first term of Eq.\,\eqref{A2}, we obtain two additional terms, i.e., $4E_c(\hat{\delta N}_x+\hat{\delta N}_q )\sigma'_z$ and $4E_cX_{\rm ZPF}(\hat{\delta N}_x+\hat{\delta N}_q )(\hat{b}^{\dagger}+\hat{b})$. The first term causes dephasing of the qubit, while the second term causes relaxation of the oscillator. When we transform $H_q$ to the qubit basis, $\sigma'_z\rightarrow(\sigma_x \cos\varphi-\sigma_z \sin\varphi$) (see below), the gate fluctuations cause both dephasing and relaxation. These fluctuations could be included in the Lindblad master equation of the total hybrid system as decay rates, and hence, could be excluded from the Hamiltonian. 
If we coherently drive the qubit at frequency $\omega_d$ and amplitude $\Omega_R$, the qubit Hamiltonian with the driving term reads as follows: 
\begin{align}
	\hat{H}_q=-\frac{\epsilon}{2}\sigma'_z-\frac{E_J}{2}\sigma'_x-\hbar g(\hat{b}^{\dagger}+\hat{b})\sigma'_z+\hbar\Omega_R \cos(\omega_dt)\sigma'_z.
\end{align}
The above Hamiltonian takes the following form, in eigenbasis of the qubit:
\begin{align} \label{A6}
	\hat{H}_q = & \frac{\hbar\omega_q}{2}\sigma_z+\hbar\Omega_R \cos(\omega_dt) (\sigma_x \cos\varphi-\sigma_z \sin\varphi) \nonumber \\
	& \hspace{1.5cm}-\hbar g(\hat{b}^{\dagger}+\hat{b})(\sigma_x \cos\varphi-\sigma_z \sin\varphi),
\end{align}
where $\tan\varphi=\epsilon/E_J$ and  $\hbar\omega_q=\sqrt{\epsilon^2+E_J^2}$.
Within the usual \textit{rotating-wave approximation} (RWA), the transverse coupling term $\hbar g(\hat{b}^{\dagger}+\hat{b})\sigma_x \cos\varphi$ could be dropped due to the significant difference in qubit and oscillator energy scales, i.e., $\omega_q \gg \Omega$. Near the charge symmetry point ($\Delta N\approx 0$ and $\sin\varphi\approx 0$), the longitudinal coupling is weak. So, we transform the transverse coupling term using the so-called  Schrieffer-Wolff transformation \cite{Hauss08} and retain the second-order coupling term, $\hbar (g^2/\omega_q)\cos^2\varphi(\hat{b}^{\dagger}+\hat{b})^2\sigma_z$.
Furthermore, due to $\omega_d\approx \omega_q$, we can drop the driving term $\hbar\Omega_R \cos(\omega_dt)\sigma_z \sin\varphi$ under the RWA.
The Hamiltonian, Eq.\,\eqref{A6} then simplifies to
\begin{align} \label{A7}
	\hat{H}_q= &\frac{\hbar\omega_q}{2}\sigma_z+\hbar\Omega_R \cos(\omega_dt)\sigma_x \cos\varphi \nonumber \\
	&+\hbar g(\hat{b}^{\dagger}+\hat{b})\sigma_z \sin\varphi+\hbar\frac{g^2}{\omega_q}\cos^2 \varphi(\hat{b}^{\dagger}+\hat{b})^2\sigma_z.
\end{align}
In the drive frame ($\omega_d$) of the qubit, and under the RWA Eq.\,\eqref{A7} could be written as follows:
\begin{align} \label{A8}
	\hat{H_q}=& -\frac{\hbar\Delta_q}{2}\sigma_z+\frac{1}{2}\hbar\Omega_R\sigma_x \cos\varphi \nonumber \\ & \hspace{-0.3cm}+\hbar g(\hat{b}^{\dagger}+\hat{b})\sigma_z \sin\varphi +\hbar\frac{g^2}{\omega_q}\cos^2 \varphi(2\hat{b}^{\dagger}\hat{b}+1)\sigma_z,
\end{align}
where $\Delta_q=\omega_d-\omega_q$. Finally, on substituting Eq.\,\eqref{A8} in Eq.\,\eqref{eq1}, we obtain the Hamiltonian of the hybrid system,described by Eq.\,\eqref{eq2}.

%-----------------------------------------
\section{Explicit Expressions for $H_{a}$ and $H_{b}$ in [Eq.\,\eqref{eq3}]} \label{AppendixB}
%%---------------------------------------
\noindent
The last two terms in Eq.\,\eqref{eq3}, results owing to the direct substitution of the transformation, $\hat{a}\rightarrow\alpha+\delta\hat{a}$ and $\hat{b}\rightarrow\beta+\delta\hat{b}$, in Eq.\,\eqref{eq2}. These terms are as follows: 
\begin{align}
H_a=&\hbar g_o(\alpha a^{\dagger}+\alpha^*a)(\beta+\beta^*) \nonumber \\ &\hspace{1cm}+\hbar\eta(a+a^{\dagger})  -\hbar\Delta_c(\alpha a^{\dagger}+\alpha^*a) 
\end{align}	
\begin{align}	
{\rm and}\hspace{1.3cm}\hspace{-0.2cm} H_b=\hbar g_o|\alpha|^2(b+b^{\dagger})+\hbar\Omega(\beta b^{\dagger}+\beta^*b).
\end{align}

\end{document}